\documentclass[preprint,showpacs,amsmath,amssymb,aps]{revtex4}
\usepackage{graphicx}
\usepackage{dcolumn}
\usepackage{bm}
\begin{document}
\title{Phase transition in liquid crystal  elastomer - 
a Monte Carlo study employing 
non-Boltzmann sampling}  
\author{
D. Jayasri ${}^{\star}$, 
N. Satyavathi ${}^{\star\star}$, 
V. S. S. Sastry ${}^{\star}$ and 
K. P. N. Murthy ${}^{\star}$}
\affiliation{
${}^{\star}$School of Physics, University of Hyderabad,
                 Hyderabad 500 046 Andhra Pradesh, India\\ 
${}^{\star\star}$ Department of Physics, Osmania University, Hyderabad
560 007, Andhra Pradesh, India 
         }
\date{\today}
\begin{abstract}
We investigate  Isotropic - Nematic transition in 
liquid crystal elastomers employing non-Boltzmann Monte Carlo techniques.
We  consider a lattice model of a liquid elastomer and 
a recently proposed Hamiltonian  which accounts for  
homogeneous/inhomogeneous interactions among liquid crystalline units,
interaction of  local nematics with  global strain, 
and with inhomogeneous  fields and stress. 
We find that when the local director is coupled strongly
to the global strain the transition is strongly first order; 
the transition  becomes weakly first order 
when the coupling becomes weaker. 
The transition temperature
decreases with decrease of coupling strength. Besides, we find 
that the 
nematic order scales nonlinearly with global strain especially 
for strong coupling and at low temperatures. 
\end{abstract}
\maketitle
\section{Introduction}  
A liquid crystal elastomer \cite{MWEMT} is a weakly cross-linked, percolating
network of long, rigid,  liquid crystalline units. It 
inherits and combines the  properties of both of its components -   
rubber elasticity of the cross-linked network and nematic and smectic 
ordering capabilities of the liquid crystalline units. As a result,  
a liquid crystal  elastomer exhibits unusual 
properties that are of interest from both basic 
\cite{basic1,basic2_strain} and applied 
science \cite{strain,artificial_muscles,tknpjcsr,ccm,artemt,stskr} 
points of views. For 
example even a small change in temperature (that causes the liquid 
crystalline units to transform from nematic to isotropic phase) can 
lead to large and reversible changes in the shape of a liquid crystal 
 elastomer 
\cite{basic2_strain,strain}. It is precisely this property that makes it 
a competitive candidate material for artificial muscles, 
see {\it e.g.} \cite{artificial_muscles,tknpjcsr}. In fact, 
since liquid crystal
elastomers respond sensitively,  not only to temperature,
 but also to 
other stimuli, like electric fields, magnetic fields, ultra violet 
radiations, gamma rays {\it etc.}, they become suitable for 
construction of actuators, detectors, micro-pumps {\it etc.}, 
see {\it e.g.}~\cite{tknpjcsr,ccm,artemt,stskr}.


We have carried out Monte Carlo simulation of a lattice
model, employing
non-Boltzmann sampling techniques. 
The transition from disordered isotropic phase at high temperatures
to ordered nematic phase at low temperatures is known to be 
weakly first order, in bulk liquid crystalline material, 
see {\it e.g.} \cite{SC}. Our simulations show that in a liquid crystal 
elastomer, when the local nematic  is coupled strongly to global elastic
order the transition is strongly first order; the transition 
becomes weak when the coupling strength decreases.
It is known that the above model exhibits strong discontinuity
for $\gamma=1$ \cite{sr} and a weak discontinuity for $\gamma=0$, see
\cite{Zhang}. Our work interpolates these two limiting cases. 
More importantly,  we 
find that the nematic order $S$, scales non-linearly
with the strain parameter  $\lambda$,  for strong coupling and  
at low temperatures. 
Our Monte 
Carlo simulations are  based 
on a modified \cite{jsm}  Wang-Landau 
implementation \cite{W_L} of entropic sampling
techniques \cite{BNJL} useful 
for studying complex systems in the neighbourhood of 
phase transition. 
In general, entropic sampling \cite{BNJL} and 
related techniques \cite{W_L} enable  
calculation of the macroscopic  properties of a closed system over
a wide  range and  at arbitrarily
fine resolution of temperatures  - all from a single entropic
ensemble that contains  microstates of all 
energies in approximately  equal proportions. More importantly
these techniques help estimate free energy and entropy, a task which is
very difficult, if not impossible, in conventional Metropolis
Markov chain Monte Carlo techniques. From the plots of free energy versus 
order parameter at various temperatures, 
we can unambiguously determine the nature of the phase transition.   
The paper is organized as follows. In section II we describe a 
lattice model of  liquid crystal elastomer and the Hamiltonian,
 proposed by
Selinger and Ratna \cite{sr}.
In section III
we describe the  non-Boltzmann 
Monte Carlo simulation techniques
employed for studying the problem. Section IV is devoted to
results and discussions. Principal outcomes of the study are 
brought out briefly in the concluding section V. 
 
\section{Lattice model of a liquid crystal elastomer}
We employ a lattice model of a liquid crystal elastomer system 
proposed by 
Selinger, Jeon  and Ratna \cite{sr}.
We consider an $L\times L\times L$ cubic lattice with lattice
sites indexed by natural numbers $i$. Each lattice site  holds a 
nematic director denoted by a unit vector $\vert u_i\rangle$. 
Each nematic director interacts 
with its six nearest neighbours and the interaction is described by
Lebwohl-Lasher potential \cite{LL}, which has head-tail flip symmetry
($\vert u_i\rangle$ and $-\vert u_i\rangle$ are equivalent).
Besides, each director is coupled to  
global elastic degree of freedom. A  model 
Hamiltonian for such an interaction between local director 
and global strain 
has been derived starting from the neo-classical theory 
of rubber elasticity \cite{MWEMT}, see  \cite{sr}.
The  Hamiltonian includes 
\begin{enumerate}
\item[(i)]
 Lebwohl-Lasher nearest neighbour interaction \cite{LL}
of the directors, placed on the lattice sites, 
\item[(ii)] the interaction of each 
local director with
\begin{enumerate}
\item[(a)] the global elastic degree of freedom and   
\item[(b)]  
an inhomogeneous  field,  
\end{enumerate}
\item[(iii)]
 an externally imposed  global stress
and 
\item[(iv)] shear modulus that couples to the strain 
and to local directors.
\end{enumerate}
It is   
given by,
\begin{widetext}
\begin{eqnarray}\label{Selinger_Ratna_Hamiltonian}
E = &-&\sum_{\langle i,j\rangle} J_{i,j}\ 
\bigg( \ \frac{3}{2}\ \langle u_i\vert u_j\rangle ^2 - \frac{1}{2}\ \bigg) 
\nonumber\\
& &  \sum_{i=1}^{L^3}
 \bigg[ \ \frac{\mu}{2}\ \bigg(\lambda^2+\frac{2}{\lambda}\bigg)
-\frac{\mu\gamma}{2}\ \bigg(\lambda^2-\frac{1}{\lambda}\bigg)\ \bigg\{
\frac{3}{2}\ \langle m\vert u_i\rangle ^2 -\frac{1}{2} \ \bigg\}-
\sigma\lambda-\langle h_i\vert u_i\rangle ^2 \ \bigg]\ .
\end{eqnarray}
\end{widetext}
In the above the first term is the Lebwohl - Lasher potential. 
The sum extends over all distinct pairs $\langle i,j\rangle$ 
of nearest neighbours. $J_{i,j} > 0$ measures the strength of nearest neighbour interaction; periodic boundary condition is imposed in all directions. 
$\mu$ is the shear modulus. For a given configuration 
(microstate ${\rm C}$) 
of
the directors, we first construct an average projection operator
given by
\begin{eqnarray}
A ({\rm C})=\frac{1}{L^3}\sum_{i=1}^{L^3} \vert u_i\rangle\langle u_i\vert\ .
\end{eqnarray}
From $A ({\rm C})$  we construct a traceless symmetric tensor,
\begin{eqnarray}
Q ( {\rm C} )=A ({\rm C})-\frac{1}{3}{\rm trace} (A)\times I\ ,
\end{eqnarray}
where $I$ denotes
a unit matrix. Let $\eta$ denote the largest eigenvalue of $Q$.
The orientational order parameter of the system when in microstate 
${\rm C}$ is given by $S=3\eta/2$.
The corresponding eigenvector is denoted by $\vert m\rangle$.
$\lambda\ge 1$ is a scalar denoting the strain parameter. We take 
$\lambda = 1+e$,
where $e$  is  
the strain along the distortion axis  taken to be  along 
the 
vector $\vert m\rangle$. 
The applied stress is denoted by $\sigma$ and $\vert h_i\rangle$ 
denotes 
an inhomogeneous field experienced by the director 
at lattice site $i$.
The coupling of the nematic to the elastic degree of freedom is 
tuned  by
the parameter $0\le \gamma\le 1$. 
The above model Hamiltonian can describe liquid crystal
elastomer with random bond ($\{ J_{i,j}\} $)  and random field
$(\{ \vert h_i\rangle \})$ disorder, 
in the presence of external stress, 
$\sigma\ne 0$. 

Selinger and Ratna \cite{sr}
considered first a homogeneous elastomer 
($J_{i,j}=1\ \forall \ \langle i,j\rangle \  
$) in the absence of local fields 
($\vert h_i\rangle =0\ \forall\  i$) with 
maximal coupling ($\gamma=1$)  and no external stress ($\sigma=0$). 
We observe   
that setting the coupling $\gamma=1$, especially at low temperatures,
is not realistic. For, such a choice would correspond to
a very steep anisotropy and would imply 
extreme elongation in one direction. On the other hand if
 $\gamma =0$, the nematic order would get decoupled from the 
global strain. Phase transitions in the liquid crystalline units will
have no effect on the shape of the elastomer.  The 
actual scenario is likely to correspond to a  choice of $\gamma $ 
between zero and unity. Accordingly, in our study 
we consider a homogeneous system
($J_{i,j}=1\ \forall \ \langle i,j\rangle$) with no external fields
($\vert h_i\rangle =0\ \forall \ i$) and  
without any applied stress ($\sigma=0$).  We  set $\mu=1$
as recommended in \cite{sr}. This corresponds to setting the 
network energy at zero strain to the highest possible energy of
the liquid crystalline units. The network is robust and deforms 
only under large 
stress.
 
\section{Monte Carlo Simulation}
A typical  Markov chain Monte Carlo simulation of the system 
would proceed as follows.
Start with an initial microstate ${\rm C}_0$ of the system
defined by a  configuration of directors $\{ \vert u_i\rangle \}$ 
and strain parameter $\lambda\ge 1$. Let $E_0$ denote the 
energy of the microstate, calculated from 
Eq.~(\ref{Selinger_Ratna_Hamiltonian}). 
Select randomly
a  director; rotate it randomly so that it points in a direction
within  a specified cone  about its current direction. 
Barker's method \cite{Barker_Watts} for selecting the trial orientation
is usually employed. 
Change the value of $\lambda$ randomly. We thus have a trial 
microstate ${\rm C}_t$ of the liquid crystal elastomer system.
Let $E_t$ be its energy. 
Accept the trial state with a 
probability 
\begin{eqnarray}
p=min\bigg( 1,\ \ \exp\ [\ -\ \beta\ \Delta E\ ]\ \bigg),
\end{eqnarray}
where $\Delta E= E_t-E_0$.
Here $\beta=1/[k_B T]$ with $T$ denoting temperature and $k_B$,
the Boltzmann constant, set to unity.
Thus, the next microstate ${\rm C}_1$ is 
either ${\rm C}_t$ with probability $p$ or ${\rm C}_0$ with probability
$1-p$. Proceed in the same fashion  and  
construct a 
Markov chain of microstates denoted by ${\rm C}_0\to {\rm  C}_1\to
{\rm C}_2\to \cdots {\rm C}_n \to \cdots$. This is called the 
Metropolis algorithm \cite{Metropolis}. 
The asymptotic $(n\to\infty)$ part of the Markov chain would 
contain microstates 
that belong to a canonical ensemble at the temperature 
chosen for the simulation.

Metropolis algorithm and it variants, see  {\it e.g.} \cite{variants},
come under the class of Boltzmann
sampling techniques. The limitations of Boltzmann sampling have  long
since been  recognized. For example it can not address
satisfactorily problems 
of super-critical slowing down near first order phase transitions,
an issue of  relevance to the simulations of liquid crystal 
elastomer systems considered in this paper.
The microstates representing the interface between the ordered and
disordered phases have intrinsically low probability of
occurrence in a closed system and hence are scarcely sampled.
Switching from one phase to the other takes a very long time due to
presence of high energy barriers when the system size is large.
As a result the relative free energies of ordered and disordered
phases can not be easily and accurately determined.

For these kinds of problems,  non-Boltzmann sampling 
provides a legitimate alternative.
Accordingly, for the simulation of lattice elastomers, we consider 
entropic sampling \cite{BNJL}, which is  
 based on the following premise.
The probability that a closed system can be found in a microstate
${\cal C}$ is given by
\begin{eqnarray}
P( {\cal C}) & = & \left[ Z(\beta)\right]^{-1}\exp[-\beta E( {\cal C})]
\end{eqnarray}
where  $E( {\cal C})$
is the energy of the microstate ${\cal C}$, 
and $Z(\beta)$ is the canonical partition function given by,
\begin{eqnarray}
Z(\beta) & = & \sum_{ {\cal C}}\exp[-\beta E( {\cal C})]\nonumber\\
         & = & \sum_E D(E)\exp[-\beta E].
\end{eqnarray}
In the above $D(E)$ is the density of states.
The probability density  for  a closed system to have  an energy $E$
is thus proportional to $D(E)\exp[-\beta E]$. Let us suppose  we want
to sample microstates in such a way that the resultant probability
density of energy is
\begin{eqnarray}
P_g (E)\propto D(E)\ [g(E)]^{-1}\label{g_ensemble}
\end{eqnarray}
where $g(E)>0\ \forall\ E $,  is a function of your choice. 
Non-Boltzmann sampling
of microstates, consistent with  Eq.~(\ref{g_ensemble}) is implemented
as follows.
Let ${\rm C}_i$ be  the current and
${\rm C}_t$ the trial microstates respectively.  
Let $E_i=E({\rm C}_i)$ and $E_t=E( {\rm C}_t)$ denote
the energy of the current and of the trial microstates respectively.
The next  entry  ${\rm C}_{i+1}$ in the Markov chain is taken  as,
${\rm C}_t$ with probability $p$ and ${\rm C}_i$ with probability $1-p$,
and $p$ is given by,
\begin{eqnarray}
p &=&
 {\rm min}\left[ 1,\frac{g(E_i)}{g(E_t)}\right].
\end{eqnarray}
It is easily verified that the above acceptance rule obeys
detailed balance  and hence we are assured that the
Markov chain constructed would converge asymptotically
to the desired $g$ ensemble. When $[g(E)]^{-1}=\exp(-\beta E)$ we
recover   conventional Boltzmann sampling implemented in the
Metropolis algorithm, described earlier. For any other choice of $g(E)$
we get non-Boltzmann sampling. However, canonical ensemble average of
a macroscopic property $O( {\rm C})$  can be obtained by
un-weighting and re-weighting of $O( {\rm C})$ for each ${\rm C}$
sampled from the $g$ ensemble; for un-weighting we
divide by $[g( E( {\rm C}))]^{-1}$ and for re-weighting we
multiply by $\exp[-\beta E( {\rm C})]$. Formally we have,
\begin{eqnarray}\label{reweighting}
\left\langle O\right\rangle &=&\frac{\sum_{ {\rm C}}
O( {\rm C}) g(E( {\rm C}))\exp[-\beta E( {\rm C})]  }
{\sum_{ {\rm C}}g(E( {\rm C}))\exp[-\beta E( {\rm C})] }.
\label{eq_reweighting}
\end{eqnarray}
The left hand side  of the above is the equilibrium value
of $O$ in a closed system at $\beta$, while in the right
side,  the summation in the numerator and in the denominator,
runs over microstates belonging to the
non-Boltzmann $g$ ensemble. The important point is that 
if we take a suitable and temperature independent $g$, then 
from a single $g$ ensemble we  can calculate  
the macroscopic properties of the system 
over a wide range of temperature. 

Entropic sampling obtains when $g(E)=D(E)$.
This choice  of $g(E)$ renders $P_g(E)$ the same for all $E$,
see Eq.~(\ref{g_ensemble}). The system does a simple
random walk on a one dimensional energy axis. 
All energy regions are visited with equal probability. The
microstates on the paths that connect ordered and disordered
phases in  a first order phase transition would  get
equally sampled. A crucial issue that needs to be addressed 
pertains to the observation that we do not know
$D(E)$  $\grave{a}$ {\it priori}.
We need a  strategy to push   $g(E)$
closer and closer to $D(E)$ iteratively. 
This we accomplish by employing Wang-Landau algorithm \cite{W_L}.
We divide the range
of energy into a  large number of equal width bins.
We denote the discrete energy version of $g(E)$ by the symbol
$\{ g_i \}$. We start with 
$g_i =1\ \forall \ i$; 
the subscript denotes energy bin index. 
We update $\{ g_i\}$
after every Monte Carlo step. Let us say the system
visits a microstate in a Monte Carlo step
and let the energy of the visited microstate
fall   in, say the $m$-th energy bin;  then $g_{m}$ is
updated to $f\times g_{m}$, where $f$ is the
Wang-Landau factor, see below. The updated
$\{ g_i\}$ becomes operative immediately for determining
the acceptance/rejection criteria of the very next trial
microstate.
We set $f=f_0 > 1 $ for the
zeroth run.
We generate a large number of microstates
employing the dynamically evolving $p$. During the run we also build 
a histogram of energy of the miocrostates visited by the system. At the end of
a run 
we check if the energy histogram 
has spanned the desired energy range and is 
reasonably flat. Flatter the histogram closer is 
$\{ g_i\}$ to $\{ D_i\}$, where $\{ D_i\}$ is discrete energy 
representation of $D(E)$. 
From one run to the next, the range of energy
spanned by the system would increase.
A run should be long enough to facilitate
the system to span an energy range  reasonably well
and to render the histogram of energy,
approximately flat.
At the end of,  say the $\nu$-th run, the
Wang-Landau factor for the next run is set as
$f=f_{\nu+1} = \sqrt{f_{\nu}}$. After several runs,
the Wang-Landau factor $f$ would be very
close to unity; this implies that there would
occur no significant change in $\{ g_i\}$
during the run. For example with the
square-root rule and $f_0=e$, we have
$f_{10}=\exp(2^{-10})=1.001$. It is clear that
$f$ decreases monotonically with increase of the
run index and asymptotically reaches unity.
Wang and Landau have recommended the square-root rule;
any other rule consistent with the above properties  of
monotonicity and asymptotic convergence to unity
would do equally well.

We  take the output $\{ g_i\}$,
which has converged reasonably to $\{ D_i\}$ (indicated by the
uniformity of energy histogram)
from the Wang-Landau Monte Carlo and carry out a single
long entropic sampling  run which generates microstates belonging
to $g$ ensemble. During this final production run we do not
update $\{ g_i\}$. By un-weighting
and re-weighting, see Eq. (\ref{eq_reweighting}), of the microstates
sampled from the $g$-ensemble during the   production run,
 we calculate the desired properties of the system as a function
of $\beta$.
This is the strategy we shall follow
in the Wang-Landau simulation of liquid crystal elastomer system,
described  in this  paper.

The usefulness of the Wang-Landau algorithm has been unambiguously
demonstrated for system with discrete energy spectrum. However
when we try to apply the technique
to systems with continuous energy, there are serious difficulties
that need careful considerations. In the present simulation we follow
the modified Wang-Landau algorithm proposed in \cite{jsm}. 
It essentially consists of treating an  entire Wang-Landau run
as a single iteration; $\{ g_i\}$ obtained at the end of a Wang-Landau
run is taken as an input for the next iteration. The convergence of
$\{ g_i\}$ to $\{ D_i\}$ is monitored by the flatness 
of the histogram of energy 
measured in each iteration. Once we get a reasonably flat histogram
we stop the iteration and start a production run with the 
final $\{g_i \}$ obtained in the last iteration. For more details of 
the simulation see \cite{jsm}.
 
\section{Results and Discussions}
We consider an homogeneous lattice liquid crystal elastomer,
with no external field and no stress: {\it i.e.} 
$J_{i,j}=1\ \forall\ \langle i,j\rangle$ and  
$\vert h_i\rangle =0\ \forall \ i$, 
$\sigma=0$ and  $\mu=1$. We consider a system with linear size  $L=6$. 
We take the initial microstate with
all $\{ \vert n_i\rangle \}$ parallel to each other and $\lambda=\lambda_0 =1$. 
Let ${\rm C}_0$ denote the initial microstate and $E_0$ 
be its energy. 
We probe the system
in an energy range $-500$ to $50$ expressed in reduced unit. 
This energy range is divided into $8000$ equal width bins.
Let $\nu$ denote the energy bin index of the initial microstate.
We set $g_i=1+\delta_{i,\nu}\ \forall\  i$.  In 
each Monte Carlo step
a director is chosen at random and its orientation is changed to a trial
orientation employing 
Barker's method \cite{Barker_Watts}. It 
essentially consists of randomly  selecting one of the three pre-fixed 
orthogonal axes 
and rotating the director about the chosen axis by 
a small angle $\Delta\theta$
sampled randomly and uniformly between $0$  and $0.02$ radians.
A trial  strain parameter, $\lambda_t$  is obtained from the 
current strain parameter $\lambda_0$ randomly following 
the prescription $\lambda_t=\lambda_0+0.01\times(\xi-0.5)$ where $\xi$
is a uniformly distributed independent random number between $0$
and $1$. These two operations give us a trial microstate ${\rm C}_t$
 of the 
lattice elastomer. The acceptance of the trial microstate is based on
entropic sampling described earlier. Once a  microstate is selected
we update $g$ as per Wang-Landau algorithm, described earlier. 
We carry out in this fashion 
one Wang-Landau run; the $g$ function at the end of a Wang-Landau
run is taken as an input for the next run. The details are 
described in \cite{jsm}. Once we get an \ \lq\ entropic function\ \rq\ 
$g$ that leads to approximately flat histogram of energy, we stop the 
iteration process and start a production run in which we obtain 
a large number microstates all belonging to  
$g$ - ensemble. From the microstates sampled from the $g$ ensemble 
we calculate all the desired macroscopic properties
of the liquid crystal elastomer system, at various temperatures
through un-weighting and re-weighting given by Eq. ({\ref{reweighting}).

Figure (1) depicts  orientational order $S$
 as a function of temperature $T$,
for various values of coupling parameter  
$\gamma=1,\ 0.8,\ 0.6,\ 0.4,\ {\rm and}\ 0$. 
For $\gamma =1$ we find that  order parameter drops  sharply
when  temperature increases. 
For smaller  $\gamma$,  the 
transition becomes less  sharp and occurs at lower 
temperature. The loss of sharpness in the transition for small $\gamma$
is also due to the small system size considered in the study.  
Figure (2) depicts the strain parameter $\lambda$
as a function of temperature for  various values of $\gamma$.
At high temperature the system is isotropic and hence irrespective 
of the coupling strength, the strain is zero {\it i.e} $\lambda=1$
for all temperatures.
When the temperature is lowered, say below $T_t$, strain develops,
see Fig. (2).
The value of $T_t$ is higher for larger 
$\gamma$. For $\gamma=1$  
the strain rises rather steeply 
with lowering of temperature. For lower values 
$\gamma$
the increase of strain with decrease 
of temperature for $T\ < \ T_t$, is not large. 
For $\gamma =0$  the orientational and 
elastic degrees of freedom are completely
de-coupled and hence  
$\lambda=1$ for all temperatures, {\it i.e.} 
no strain develops
even when temperature is lowered and nematic order sets in. 
These results are consistent with 
experimental observations.

Figure (3) depicts  specific heat, calculated from 
the fluctuations of energy, as a function of temperature for various
values of $\gamma$. For $\gamma =1$, we find that the specific heat shows
a sharp maximum at the transition temperature. For lower values of $\gamma$
the transition occurs at lower temperatures. This is in agreement 
with the mean-field arguments and Uchida \cite{N_Uchida}; 
also the peak is broader 
at lower $\gamma$.  

Figure (4) depicts the transition temperature as a function of 
$\gamma$. The transition temperature is taken as the value of $T$
at which the specific heat shows a maximum. As $\gamma$ increases  
the transition temperature increases slowly initially; when $\gamma $
increases beyond $0.6$ the transition temperature increases rather steeply. 

As seen from Figs. (1) and (2),  both   
$S$ and $\lambda$ increase with decrease 
of temperature. To see the 
nature of their correlation we have plotted in Fig. (5), $\lambda$ versus
$S$ for various values of $\gamma$. For $\gamma=0$ the strain 
is zero ($\lambda =1$) and is independent of $S$ as expected since 
in this case the orientational and elastic degrees of freedom are 
uncorrelated.  For small values of $\gamma$ the strain 
parameter $\lambda$ scales 
linearly with $S$ over the full range of temperature. 
For $\gamma =1$, the scaling is linear  
for $S\le 0.6$ and $\lambda \le 1.25$ which correspond to 
temperature greater than about $0.8$. This is consistent 
with the results of earlier simulation~\cite{sr}  which showed linear 
scaling between
$\lambda$ and $S$. However, we find that for 
lower temperatures the scaling of $\lambda$ with $S$ is nonlinear.
The strain increases steeply to large values as the orientational 
order increases and attains its maximum value of unity.  

One of the advantages 
of multicanonical Monte Carlo techniques is that we
can calculate density of states up to a normalization.
The  array $\{ g_i \}$ would give an approximate
estimate of density of states if the corresponding histogram of energy
of sampled microstates is relatively flat. Fig. (6) depicts 
$g$ function and the corresponding energy histogram.  
The energy histogram is relatively flat. From the $g$ function
we can calculate the microcanonical entropy and free energy 
up to an  additive constant. 
Note that entropy or free energy calculations are very difficult 
if not impossible, from conventional Metropolis Monte Carlo simulations. 
We show in Fig. (7)
variation of (relative) free energy with energy  
for $\gamma =1$ at three  temperatures bracketing the transition 
point. We see clearly that the transition is first order and 
strong. Figure (8) depicts free energy versus $E$ for $\gamma=0.8$.
The transition is still first order but is relatively weak.
For smaller values of $\gamma$ the transition weakens further.
For $\gamma$ less than $0.4$ or so, the free energy barrier is not discernible.
 
\section{Conclusions}
We have reported in this paper results on phase transition
in liquid crystal elastomers employing multicanonical
Monte Carlo simulations. We have considered the nature of transition
for various strengths of coupling of the local nematic director
to a global strain. We find that at maximal coupling,  the transition
is strongly first order. As the coupling becomes weaker, the transition
becomes weakly first order. 
We find that the scaling of nematic order with  strain is 
non-linear especially when the local nematic directors are strongly
coupled to global strain and at low temperatures.\\  

\begin{acknowledgements}
This work has been carried out under the 
Board of Research in Nuclear Sciences (India) project No. 
2005/37/28/BRNS/1820. The Monte Carlo simulations reported 
in this paper were carried out at the Centre for 
Modeling and Simulation (CMSD) of the Hyderabad University. 
We are thankful the anonymous referees for their criticisms and 
suggestions.
\end{acknowledgements}


\begin{center}
\begin{figure*} 
\includegraphics[height=200mm,width=180mm]{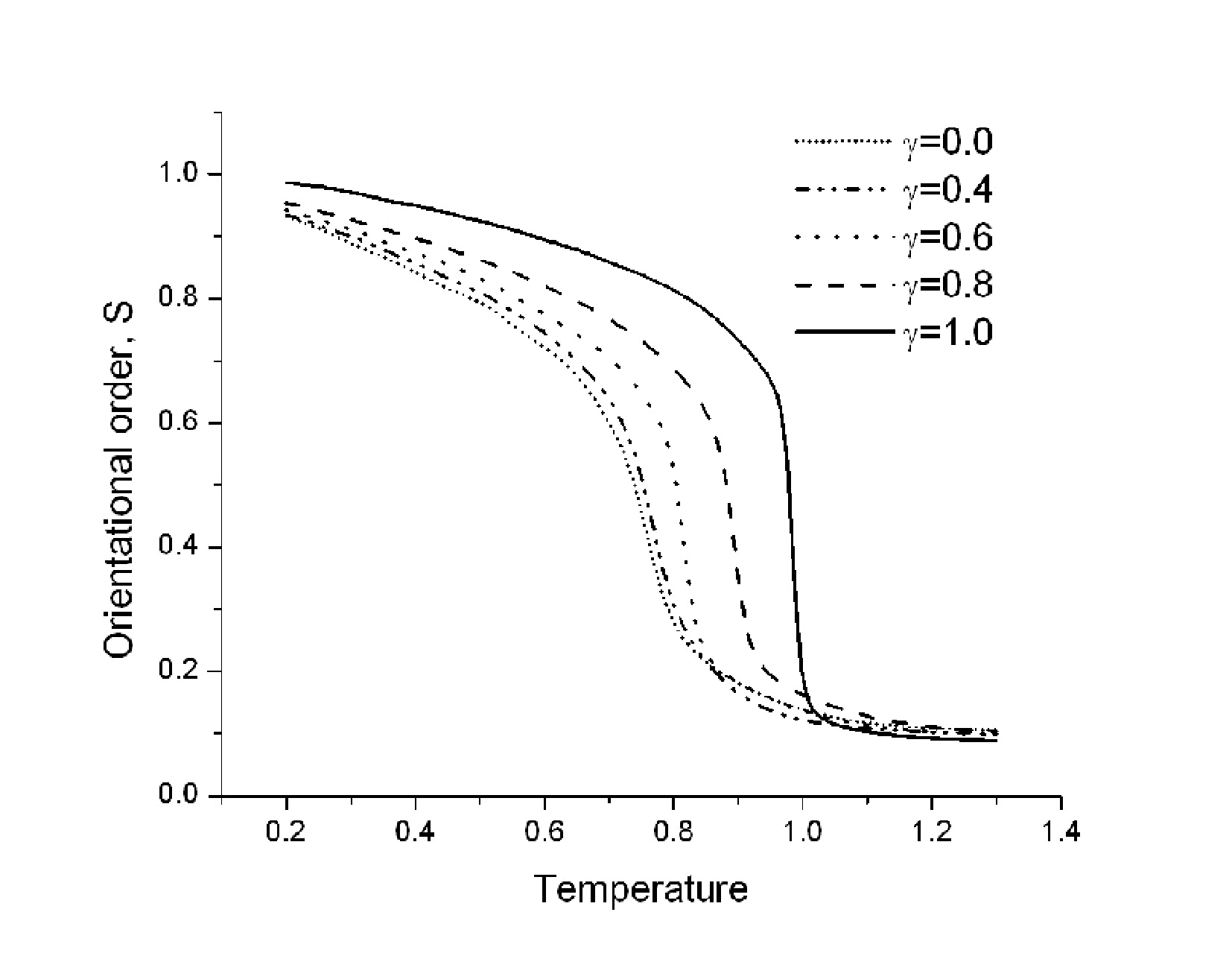}
\caption{Orientational ordre parameter $S$ versus temperature for
various values of the the coupling parameter $\gamma$.}
\end{figure*}
\end{center}

\begin{center}
\begin{figure*} 
\includegraphics[height=200mm,width=180mm]{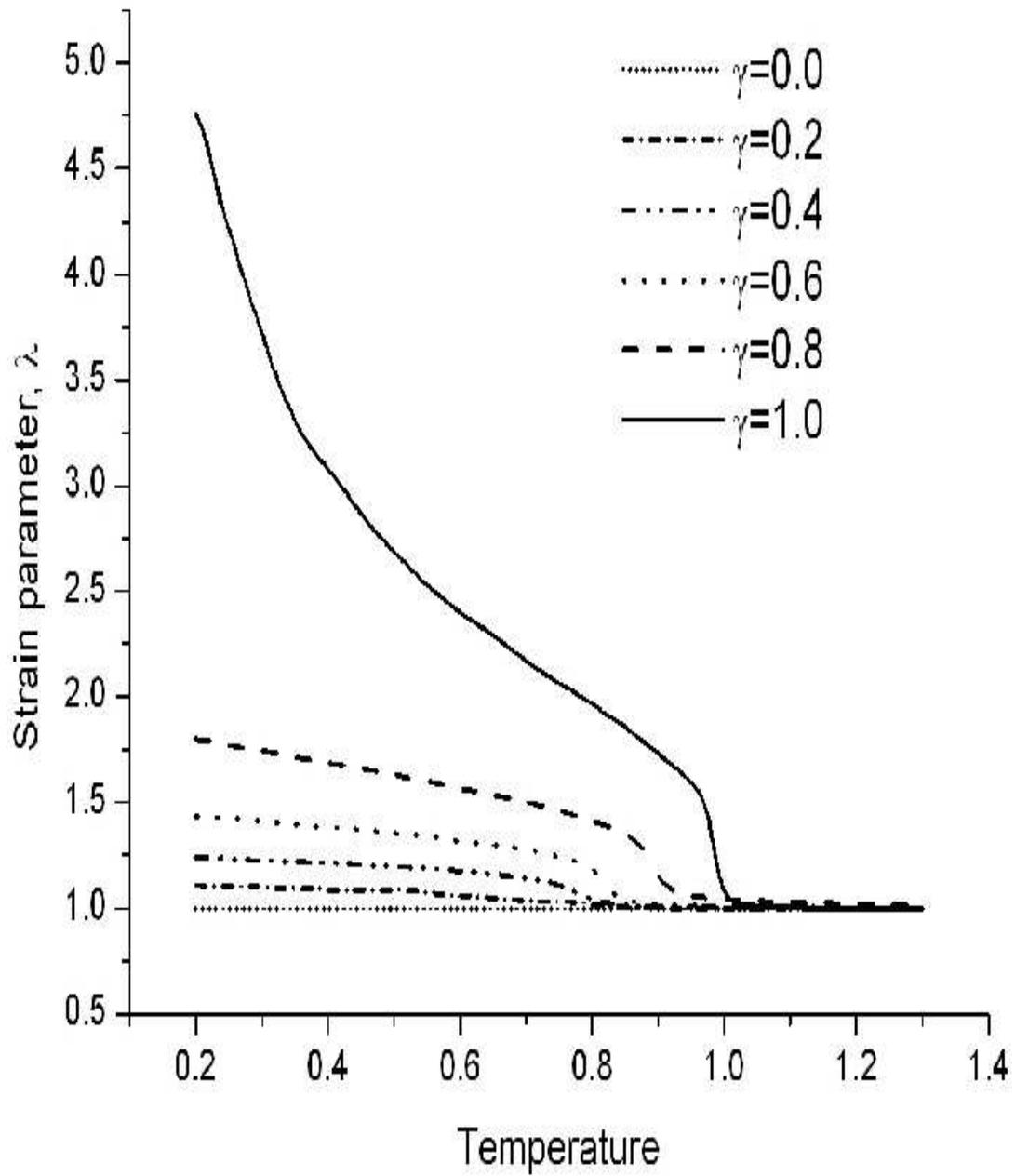}
\caption{Strain parameter $\lambda$ versus temperature
for various values of the the coupling parameter $\gamma$.}
\end{figure*}
\end{center}

\begin{center}
\begin{figure*} 
\includegraphics[height=200mm,width=180mm]{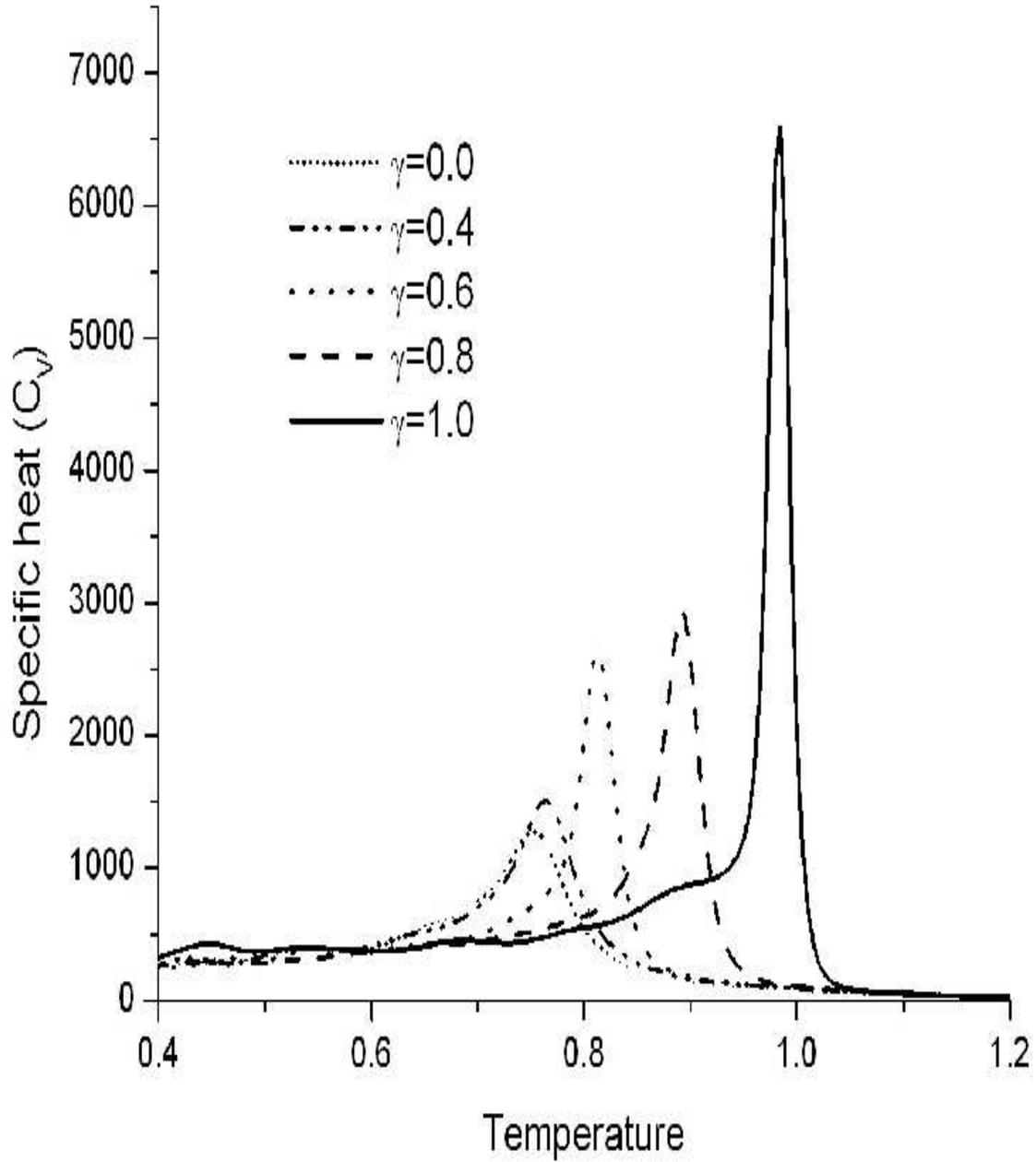}
\caption{Specific heat ${\rm C}_V$ 
versus temperature for various values of 
the coupling parameter $\gamma$; ${\rm C}_V$ has been obtained from
the energy fluctuations.
}
\end{figure*}
\end{center}

\begin{center}
\begin{figure*} 
\includegraphics[height=200mm,width=180mm]{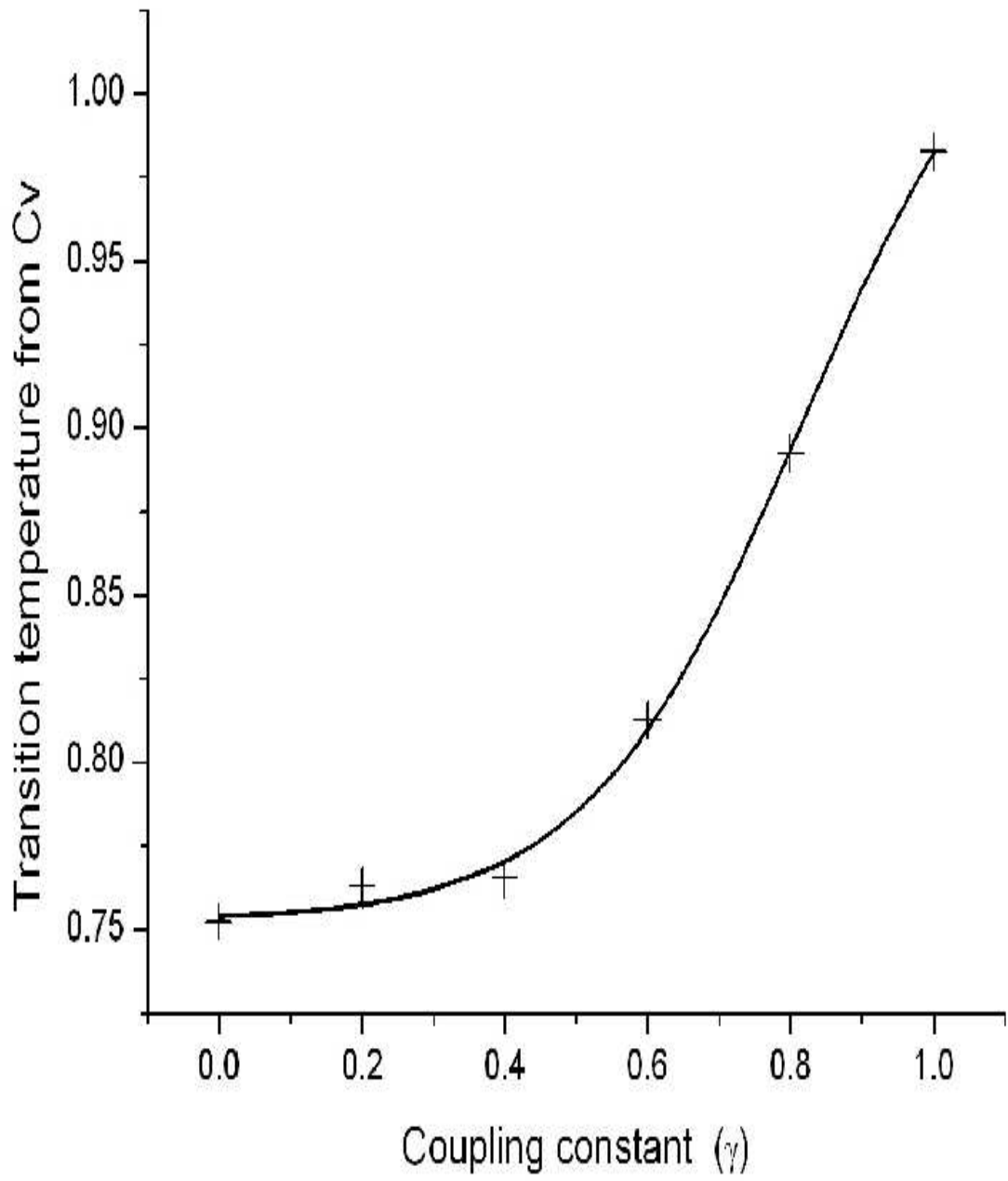}
\caption{Transition temperature versus the 
coupling parameter $\gamma$; the temperature at which 
the specific heat exhibits peak is taken as the transition point. 
}
\end{figure*}
\end{center}

\begin{center}
\begin{figure*} 
\includegraphics[height=200mm,width=180mm]{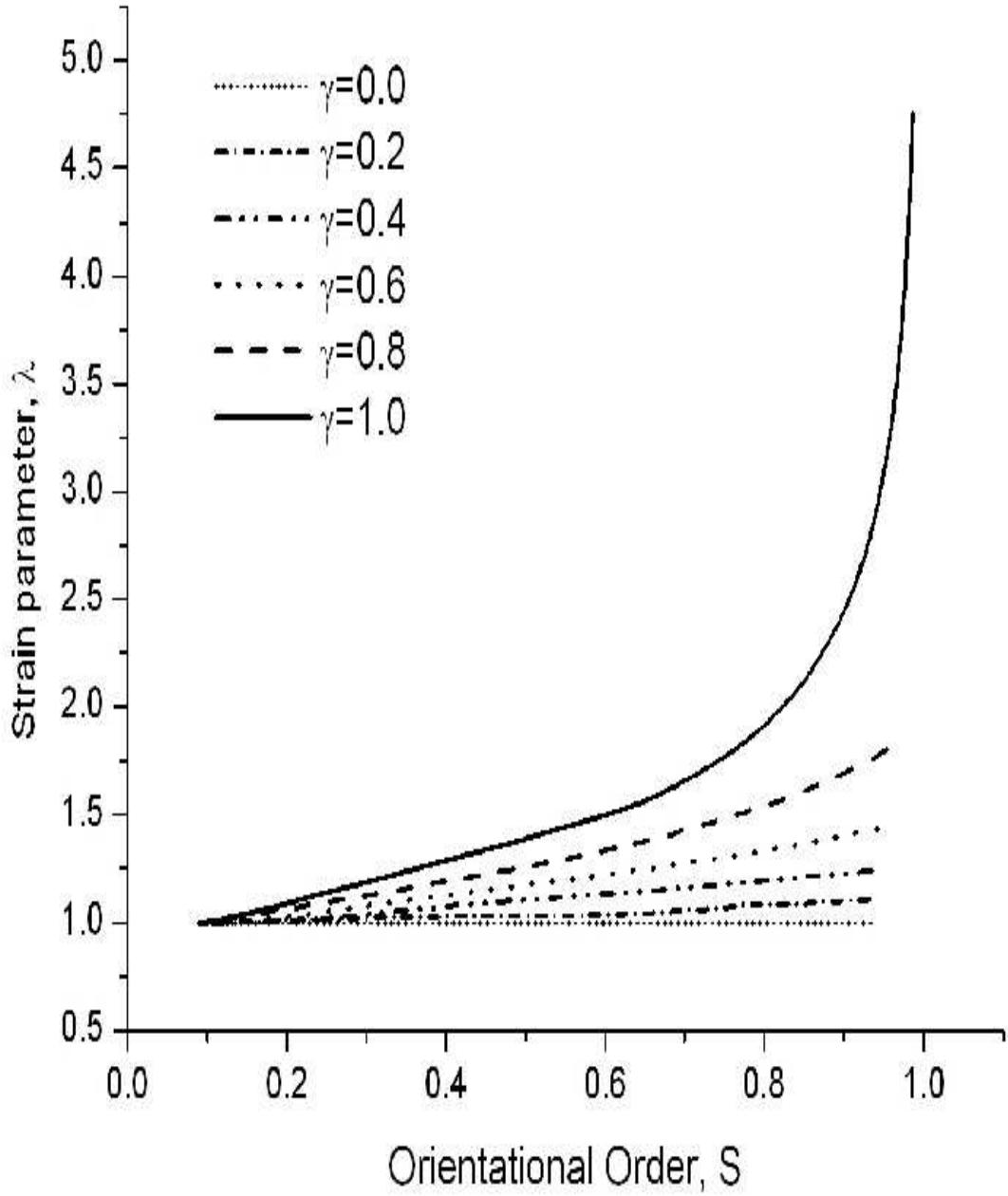}
\caption{Scaling of Strain ($\lambda$) with orientational order 
($S$)
}
\end{figure*}
\end{center}

\begin{center}
\begin{figure*} 
\includegraphics[height=200mm,width=180mm]{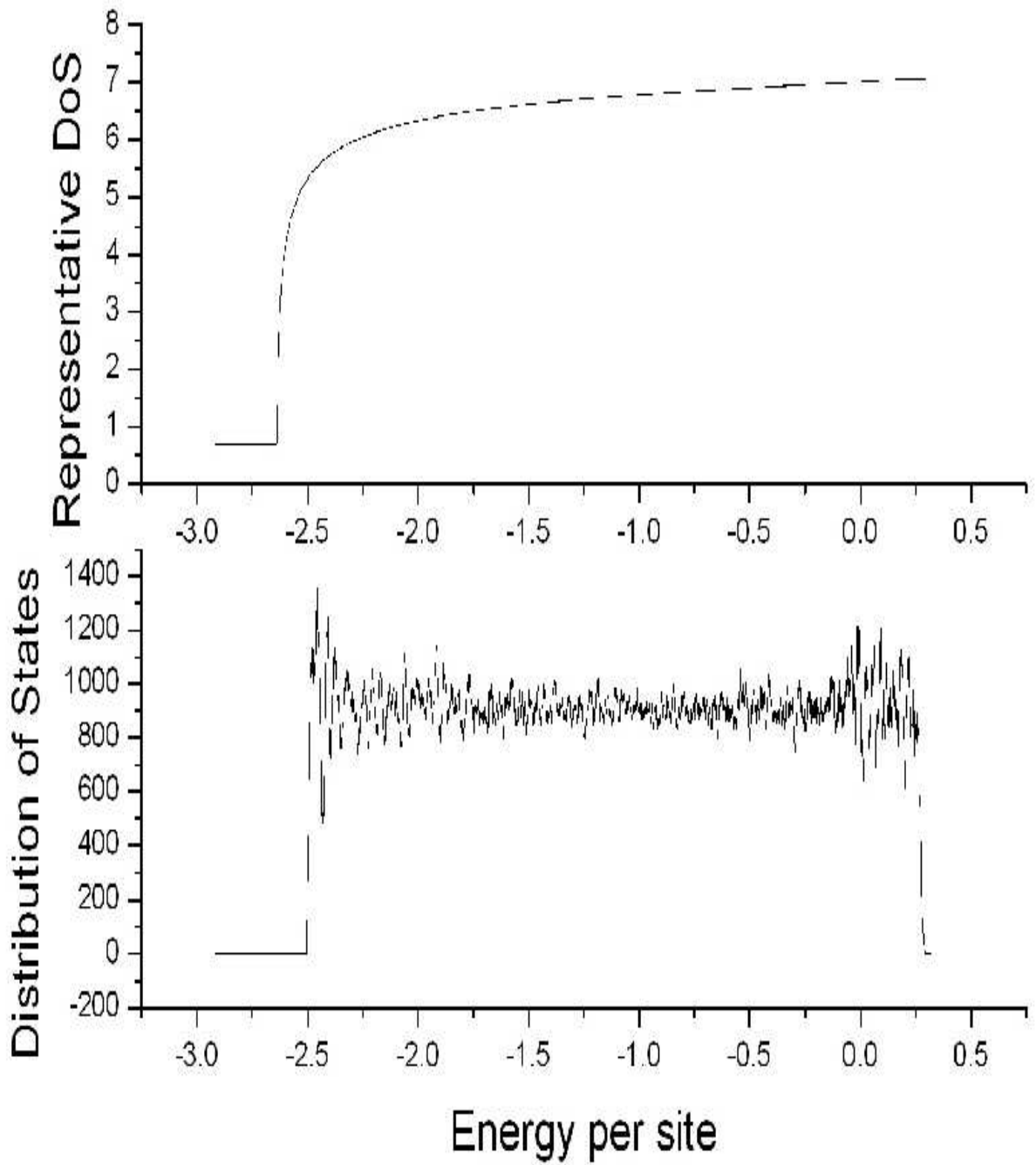}
\caption{(Top) The $g$ function that approximates the 
density of states $D$; logarithm of $g$ gives the microcanonical entropy;
what is plotted is logarithm of entropy as a function of ewnergy. 
(Bottom) The histogram of energy of mictrostates 
sampled during the production run employing entroping sampling 
with acceptance determined by the density of states depicted 
in the top. 
}
\end{figure*}
\end{center}

\begin{center}
\begin{figure*} 
\includegraphics[height=200mm,width=180mm]{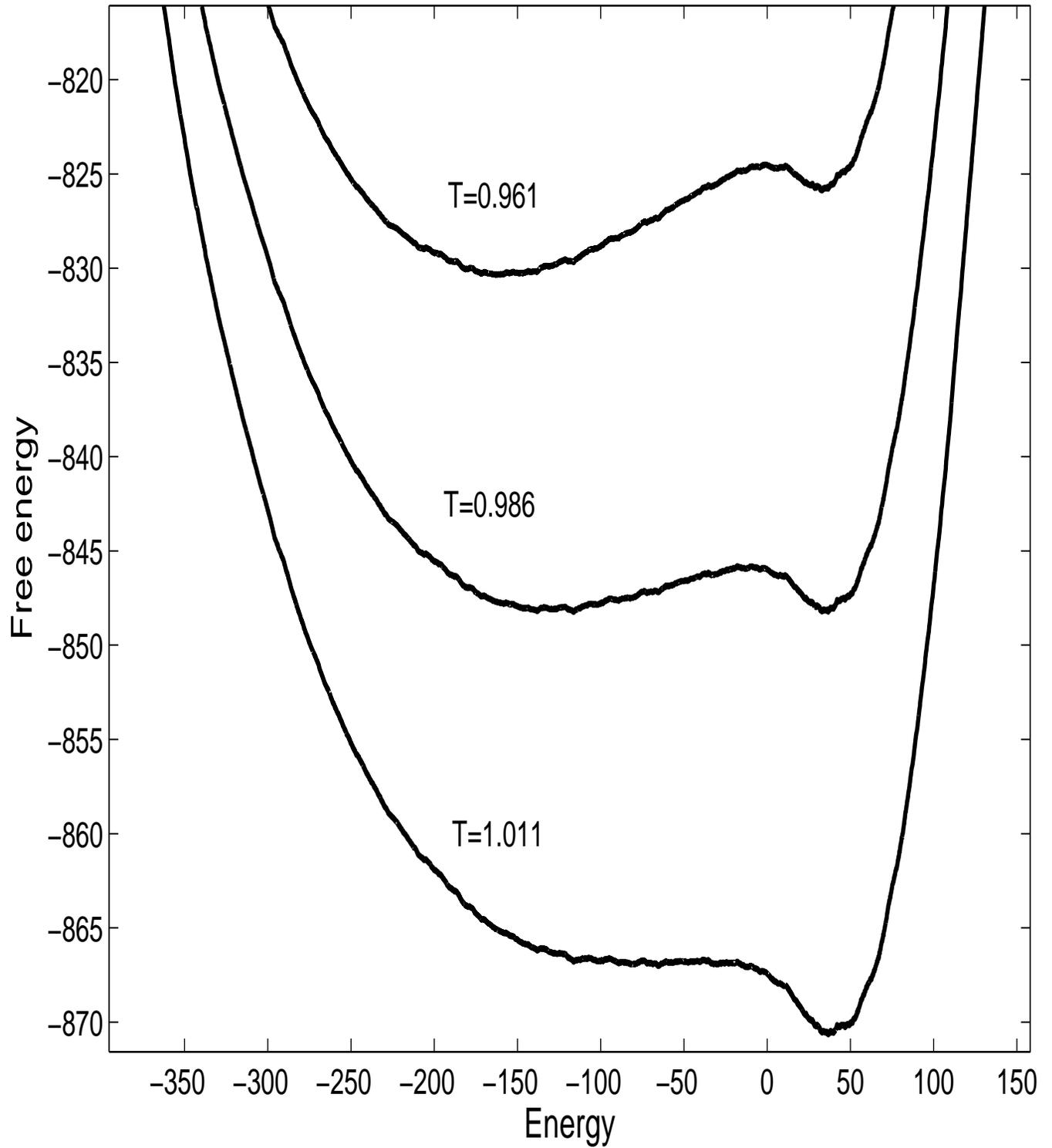}
\caption{Free energy versus energy for temperatures above,
below and at the transition point for the system with 
$\gamma=1.0$. It is clear that the transition is first order.
}
\end{figure*}
\end{center}

\begin{center}
\begin{figure*} 
\includegraphics[height=200mm,width=180mm]{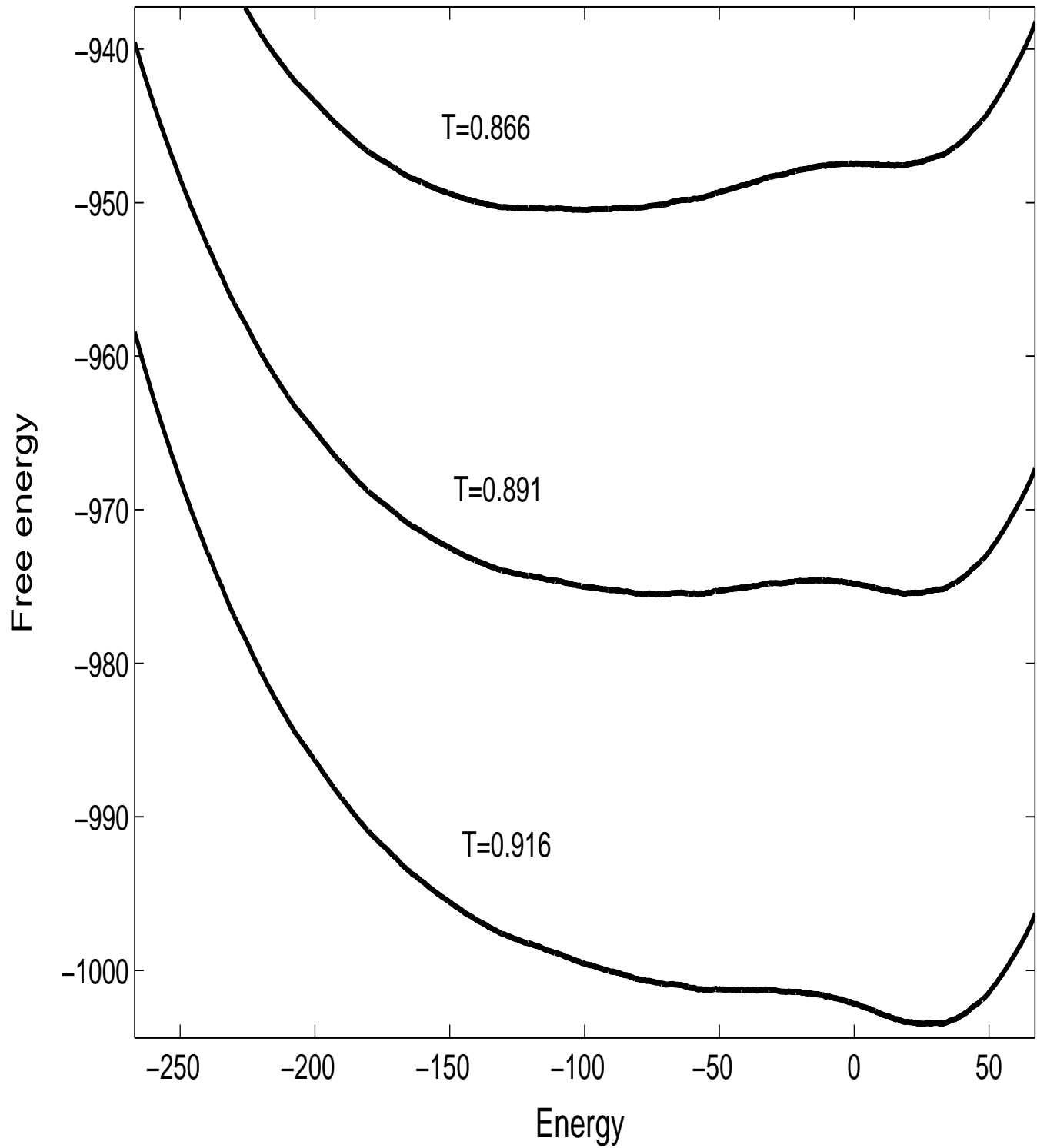}
\caption{Free energy versus energy for temperatures above, 
below and at the transition point for the system with 
$\gamma=0.8$. The barrier height is small; the transition is 
still first order though weak.
}
\end{figure*}
\end{center}
\end{document}